\documentclass[sigconf, table]{acmart}
\renewcommand\footnotetextcopyrightpermission[1]{} 
\usepackage[utf8]{inputenc}
\usepackage{algorithmic}
\usepackage{algorithm}
\usepackage{amsfonts}
\usepackage{stmaryrd}
\usepackage{amsmath}
\usepackage{bm}
\usepackage{isomath}
\usepackage{subcaption}
\usepackage{hyperref}
\DeclareMathOperator*{\argmax}{arg\,max}
\DeclareMathOperator*{\argmin}{arg\,min}

\newtheorem{theorem}{Theorem}

\copyrightyear{2022}
\acmYear{2022}
\setcopyright{acmcopyright}\acmConference[MSWiM '22]{Proceedings of the International Conference on Modeling Analysis and Simulation of Wireless and Mobile Systems}{October 24--28, 2022}{Montreal, QC, Canada}
\acmBooktitle{Proceedings of the International Conference on Modeling Analysis and Simulation of Wireless and Mobile Systems (MSWiM '22), October 24--28, 2022, Montreal, QC, Canada}
\acmPrice{15.00}
\acmDOI{10.1145/3551659.3559050}
\acmISBN{978-1-4503-9479-6/22/10}

\usepackage{tikz}
\def\checkmark{\tikz\fill[scale=0.4](0,.25) -- (.25,0) -- (.7,.6) -- (.25,.15) -- cycle;} 

\begin{document}

\title[INSPIRE: Distributed BO for ImproviNg SPatIal REuse in Dense WLANs]{INSPIRE: Distributed Bayesian Optimization for ImproviNg SPatIal REuse in Dense WLANs}

\author{Anthony Bardou}
\email{anthony.bardou@ens-lyon.fr}
\orcid{0000-0003-3238-0274}
\affiliation{%
  \institution{Univ Lyon, ENS de Lyon\\Université Claude Bernard Lyon 1, CNRS, LIP}
  \streetaddress{46 Allée d'Italie}
  \city{Lyon}
  \country{France}
  \postcode{69007}
}

\author{Thomas Begin}
\email{thomas.begin@ens-lyon.fr}
\affiliation{%
  \institution{Univ Lyon, ENS de Lyon\\Université Claude Bernard Lyon 1, CNRS, LIP}
  \streetaddress{46 Allée d'Italie}
  \city{Lyon}
  \country{France}
  \postcode{69007}
}

\begin{abstract}
  WLANs, which have overtaken wired networks to become the primary means of connecting devices to the Internet, are prone to performance issues due to the scarcity of space in the radio spectrum. 
As a response, IEEE 802.11ax and subsequent amendments aim at increasing the spatial reuse of a radio channel by allowing the dynamic update of two key parameters in wireless transmission: the transmission power (\texttt{TX\_POWER}) and the sensitivity threshold (\texttt{OBSS\_PD}). In this paper, we present \texttt{INSPIRE}, a distributed online learning solution performing local Bayesian optimizations based on Gaussian processes to improve the spatial reuse in WLANs.
\texttt{INSPIRE} makes no explicit assumptions about the topology of WLANs and favors altruistic behaviors of the access points, leading them to find adequate configurations of their \texttt{TX\_POWER} and \texttt{OBSS\_PD} parameters for the ``greater good'' of the WLANs. 
We demonstrate the superiority of \texttt{INSPIRE} over other state-of-the-art strategies using the ns-3 simulator and two examples inspired by real-life deployments of dense WLANs. 
Our results show that, in only a few seconds, \texttt{INSPIRE} is able to drastically increase the quality of service of operational WLANs by improving their fairness and throughput.
\end{abstract}

\ccsdesc[500]{Networks~Wireless local area networks}
\ccsdesc[500]{Theory of computation~Bayesian analysis}
\ccsdesc[500]{Theory of computation~Reinforcement learning}
\ccsdesc[500]{Theory of computation~Gaussian processes}
\ccsdesc[500]{Theory of computation~Distributed algorithms}
\ccsdesc[500]{Networks~Network performance analysis}

\keywords{Machine Learning; Gaussian Process; Spatial Reuse; IEEE 802.11; Power Control}

\newcommand{\TB}[1]{\textcolor{orange}{\em  #1}}
\newcommand{\Bardou}[1]{\textcolor{purple}{\em  #1}}

\maketitle

\keywords{Machine Learning, Spatial Reuse, IEEE 802.11, Power Control}

\section{Introduction}

Since their introduction in the late 1990s, WLANs (Wireless Local Area Networks) have rapidly overtaken wired networks to become the primary means of connecting devices to the Internet. 
According to Cisco \cite{cisco2019wp}, they will account for 57\% of the Internet traffic in 2022, compared to 22\% and 21\% for mobile and wired networks, respectively. 
The current WLAN architecture is defined by the IEEE standard 802.11 (commercially known as Wi-Fi). APs (Access Points) are the centerpiece of this setup, serving as relays for wireless devices; we refer to the latter as STAs (Stations) throughout this paper. 
Typically, each AP is equipped with a wired interface giving access to the LAN and then the Internet as well as a wireless interface providing connectivity to nearby STAs through radio communications.

Space on the radio spectrum is a scarce resource as it is often shared by multiple WLANs. 
The radio bands used by the IEEE 802.11 standard (currently 2.4 and 5 GHz, soon to be joined by 6 GHz) are divided into channels.  
Different APs can then be assigned to different, orthogonal channels enabling the APs to transmit at the same time without interfering with each other. 
Equally important thanks to the limited radio range of radio waves, APs configured on the same radio channel can transmit concurrently provided that they are sufficiently far away from each other. 
This ability was central to the success of WLANs and it is commonly known as the spatial reuse of radio channels.

However, the spatial reuse of radio channels as performed by today’s WLANs may be reaching its limit. 
This is particularly true in places where WLAN deployments are very dense, such as offices, shopping malls and train stations. 
This is because, in these areas the distance between APs is  small, so that an AP is more likely to be blocked by the transmissions of one or several nearby APs operating on the same channel. 
This will in turn  take a hefty toll on the WLANs' performance.

\begin{figure*}[t]
    \centering
    \begin{subfigure}{0.33\textwidth}
        \centering
        \includegraphics[width=\linewidth]{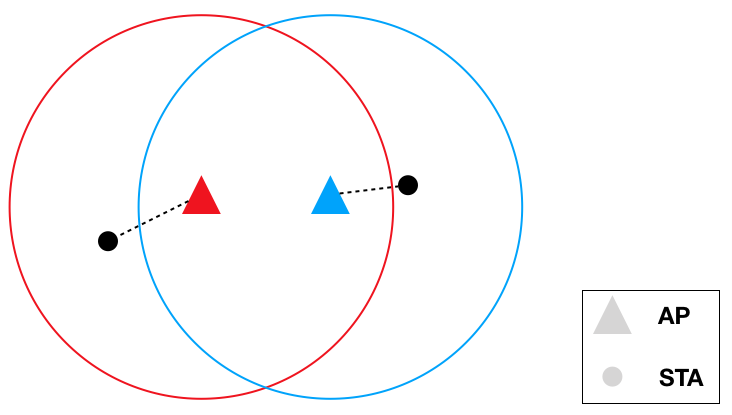}
        \caption{With the default configuration of \texttt{TX\_PWR} and \texttt{OBSS\_PD}, the two APs are within each other’s detection range so that they cannot transmit simultaneously.} 
        \label{fig:toya}
    \end{subfigure}
    \hspace{1.5cm}
    \begin{subfigure}{0.33\textwidth}
        \centering
        \includegraphics[width=\textwidth]{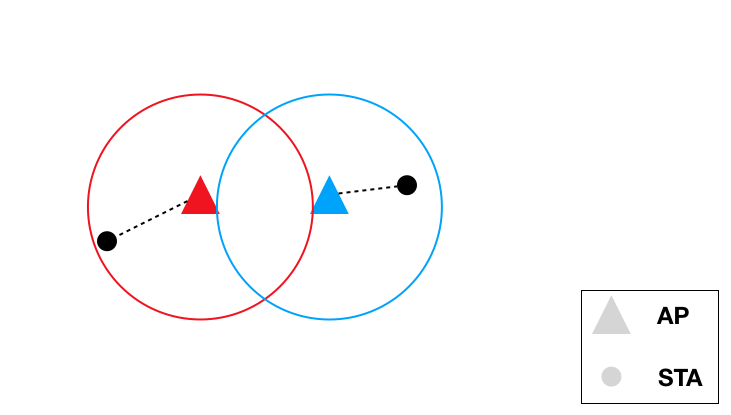}
        \caption{The value of \texttt{TX\_PWR} is reduced on both AP so that they do not belong to each other’s detection range. Under this configuration, concurrent transmissions from the two AP may occur at the same time.}
        \label{fig:toyb}
    \end{subfigure}
    \caption{Adequately configuring the \texttt{TX\_PWR} parameter of APs can significantly improve the spatial reuse of radio channels in WLANs. Note that concurrent transmissions of the two APs could also be attained by increasing \texttt{OBSS\_PD} at each AP. While similar, reducing \texttt{TX\_PWR} and increasing \texttt{OBSS\_PD} may affect the WLANs' performance differently (see Table 2 of \cite{sr_mab_2} for more details).}
    \label{fig:toy}
\end{figure*}

A solution to this issue can be found in the 2021 amendment to 802.11 known as 802.11ax \cite{802.11-2021}, which enables the dynamic configuration of two key parameters at each AP: \texttt{TX\_PWR} and \texttt{OBSS\_PD}. 
The former parameter specifies the power level (in dBm) at which the AP transmits its data. 
The latter parameter defines the sensitivity threshold (in dBm). 
If the energy received is below this level, this indicates to the AP that the radio channel is clear and thus available for transmission. Otherwise, the AP must defer its transmissions. 
While prior amendments to 802.11 held these \texttt{TX\_PWR} and \texttt{OBSS\_PD} parameters constant (typically 20 dBm and -82 dBm respectively), 802.11ax has made them dynamic with their values spanning from 1 to 21 dBm for the former and from -82 to -62 dBm for the latter.  
Adjusting the configurations of \texttt{TX\_PWR} and \texttt{OBSS\_PD}  can help overcome the limitations of spatial reuse in dense environments by allowing APs that are close to each other to transmit on the same channel. 
Figure~\ref{fig:toy} depicts a simple example of two APs operating on the same radio channel and illustrates how different configurations of the \texttt{TX\_PWR} and \texttt{OBSS\_PD} parameters can lead to different performance. 

Despite the potential of 802.11ax to improve the spatial reuse of radio channels, finding an adequate configuration of \texttt{TX\_PWR} and \texttt{OBSS\_PD} for the APs in a WLAN is a complex problem. 
First, an adequate configuration is very topology-specific. 
In other words, knowing a suitable configuration for a given scenario is of no value for another scenario. 
Second, a distributed solution would be more appreciated than a centralized solution. 
Not only does this avoid the search in an otherwise very high dimensional space but this avoids the assumption of having a centralized entity (e.g., a controller) deciding the configurations of all APs. 
This assumption is acceptable if all the interfering APs belong to the same WLAN but unrealistic if they belong to concurrent WLANs. Third, forecasting the performance of WLANs with an analytical model that can subsequently help establish ``optimal’’ configurations is difficult. 
The degree of details in the models will be either too coarse and thus inapplicable, or adequate but unscalable when scenarios involve multiple APs and STAs. 
A solution to this is that APs can apply new configurations of their parameters, measure the effect of these changes on their performance, and exchange their experience with surrounding APs. 
This paves the way for the use of online and reinforcement learning techniques in a distributed manner.

In a previous work \cite{mab_mswim}, we introduced a centralized solution based on the sampling of Gaussian mixtures and Thompson sampling to improve the spatial reuse of WLANs by adequately configuring \texttt{TX\_PWR} and \texttt{OBSS\_PD} parameters at each AP. In this paper, we present a distinct contribution named \texttt{INSPIRE}, which is a distributed online learning solution performing local Bayesian optimizations based on GPs (Gaussian Processes)
to address the issue of improving the spatial reuse of WLANs. 
\texttt{INSPIRE} makes no explicit assumptions about the topology of WLANs or the radio environments and thus can apply to any WLANs. 
Additionally, it can operate even when the APs to be configured belong to different concurrent WLANs. 
The contributions of this paper are as follows:
\begin{itemize}
\item We demonstrate the ability of GPs at approximating the reward function, which reflects the performance of WLANs, and at exploring efficient AP configurations; 
\item We establish the superiority of a divide-and-conquer approach to handle the complex problem of setting the \texttt{TX\_PWR} and \texttt{OBSS\_PD} parameters at each AP; 
\item We introduce \texttt{INSPIRE} a distributed solution that lets the APs of concurrent WLANs automatically adapt their internal parameters’ setting in their own interest as well as in the interest of obtaining a more efficient spatial reuse of radio channels;
\item We evaluate the efficiency of \texttt{INSPIRE} on real-life inspired case studies using a detailed network discrete-event simulator and compare its performance with several state-of-the-art solutions (centralized or distributed).
\end{itemize}

The remainder of this paper is organized as follows. 
The next section discusses the related work. 
Section~\ref{sec:sol} describes the proposed strategy and the main theoretical results obtained to address the issue of spatial reuse of a radio channel in WLANs. 
Its performance are evaluated in Section~\ref{sec:num_res}.
Section~\ref{sec:conc} concludes this paper. 

\section{Related work} \label{sec:soa}

The release of the IEEE 802.11ax amendment \cite{802.11-2020} in late 2021 marks a new era for the spatial reuse of radio channels of WLANs:  Nodes can dynamically adjust their transmission power (\texttt{TX\_PWR}) and sensitivity threshold (\texttt{OBSS\_PD}) parameters. 
For a detailed explanation of how this new feature is implemented, we refer the interested reader to~\cite{Wilhem2021}, which also provides simple scenarios to illustrate its potential benefits.

Years before IEEE released the 802.11ax amendment, the idea of dynamically updating \texttt{TX\_PWR} and \texttt{OBSS\_PD} has been explored by some researchers. 
The pioneering work of \cite{Zhu2004} presents an analytical model that, based on the current radio channel conditions, dynamically configures \texttt{OBSS\_PD} on each node of a Wi-Fi-based mesh network. 
Concurrently, \cite{Kim2004} established that adapting \texttt{TX\_PWR} can lead to increased throughput and reduced energy consumption. 
More recently, in 2020, \cite{Qiu2020} casts the issues of positioning the APs of a WLAN and choosing their \texttt{TX\_PWR} as an optimization problem. 
The authors provides a solution to this problem that delivers a static configuration of \texttt{TX\_PWR} for a WLAN. 
But their solution does no account for the number of STAs nor the type of traffic in the WLAN.

The difficulty of accurately modeling the dependency between the configuration parameters of a large WLAN and its performance is a strong hurdle to the development of spatial reuse strategies based on analytical models. 
As a result, most of the proposed strategies are data-driven. 
Adaptive by construction, they seem to constitute promising candidates in the search for configurations that improve the spatial reuse of a radio channel and the performance of WLANs.

Machine learning (ML) techniques are natural candidates for addressing problems requiring a data-driven approach, and the spatial reuse problem is no exception. 
\cite{fsc} addresses the problem of configuring \texttt{TX\_PWR} and \texttt{OBSS\_PD} with a two-scale solution using artificial neural networks (ANN). 
In their strategy, STAs and APs first adjust their value of \texttt{OBSS\_PD} to minimize interference. 
Then, an ANN, which was trained offline through simulation, is used to increase the fairness between STAs in terms of attained throughput. 
However, given the vast diversity of WLAN topologies, the offline learning of the ANN appears as a clear limitation to the generalization of this strategy.  
An online learning procedure is proposed by \cite{mab_mswim}, which uses reinforcement learning and more precisely the Multi-Armed Bandit (MAB) framework to find the optimal configuration of \texttt{TX\_PWR} and \texttt{OBSS\_PD} in a WLAN. 
The approach comprises two agents with one sampling promising configurations through a multivariate normal distribution, and the other identifying the best configuration among those already sampled with Thompson sampling and Normal-Gamma priors. 
These two ML solutions \cite{fsc,mab_mswim} were tested on the network simulator ns-3 and lead to significant WLAN improvements. 
However, in order to perform their optimization, they both assume the presence of a central controller that has access and control over all the APs in the WLANs. 
By construction, these approaches are centralized, and hence cannot be applied to cases where concurrent WLANs managed by different owners interfere with others.

Distributed approaches are undisputedly better fit than centralized approaches to handle cases with a set of concurrent WLANs. 
\cite{dsc} introduces a distributed algorithm named Dynamic Sensitive Control which is run on every STAs of a WLAN. 
In short, each STA tries to dynamically reduce its value of \texttt{OBSS\_PD} to favor concurrent transmissions while keeping it high enough to ensure a high quality signal reception. 
Similarly, \cite{lsr} proposes Link-aware Spatial Reuse (LSR), a distributed algorithm designed for the APs. 
In LSR, each AP chooses another AP, which is allowed to transmit concurrently, and then prescribes a value of \texttt{TX\_PWR} for the selected AP.
These two algorithms rely on a single measurement metric reflecting the quality of the received signal, namely the Received Signal Strength, to choose the nodes configuration. 
More recently, strategies using distributed MAB approaches have been proposed \cite{sr_mab, sr_mab_2}. 
They both use Thompson sampling with Gaussian priors to find the best couple of \texttt{TX\_PWR} and \texttt{OBSS\_PD} at each AP. 
In~\cite{sr_mab}, each AP seeks to maximize the throughput of its associated STAs.
On the other hand, in~\cite{sr_mab_2}, the authors assume that every AP has access to the performance of all other APs in the WLAN; then each AP attempts to maximize a global reward that takes into account the performance of all the other nodes. 
Both strategies \cite{sr_mab, sr_mab_2} solutions were evaluated in a self-made simulator with simple random scenarios.

Table \ref{tab:soa} summarizes the main characteristics of the data-driven strategies discussed above. 
It shows that, out of the six considered strategies, two (namely \cite{dsc,fsc}) only focus on the configuration of  the \texttt{OBSS\_PD} parameter (keeping the \texttt{TX\_PWR} parameter fixed). 
To help in the comparison of the different strategies, we introduce two concepts: ``pull area'' and ``push area''. 
The pull area indicates the area on which each node is assumed to obtain information (this typically includes parameter configurations and performance measurements). 
Depending on the strategy being considered, the pull area can include just the node itself, the surrounding nodes, or the whole set of nodes in the WLANs. 
The push area designates the area which each AP can influence typically through the prescription of parameter configurations. 
In the case of centralized strategies (e.g., \cite{fsc, mab_mswim}), the pull and push areas naturally cover the whole set of APs. 
We distinguish partially distributed strategies (e.g., \cite{sr_mab_2}) wherein either the pull or push area includes the whole set of APs with fully distributed strategies (e.g., \cite{dsc, lsr, sr_mab}) in which both the pull and push areas differ from the whole set of APs. 
We observe in Table~\ref{tab:soa} that only three out of the six state-of-the-art strategies can be considered as fully distributed. 
The last four columns of Table~\ref{tab:soa} pertain to the performance evaluation used to validate each of these strategies. 
It appears that most strategies were evaluated without considering the dynamical selection of the Modulation Coding Scheme (MCS) for the speed of the wireless links, nor bidirectional (with upstream and downstream) traffic. 
This can be seen as a strong limitation since this overlooks some associated trade-offs. 
For instance, increasing the value of \texttt{TX\_PWR} certainly enables the communication to operate with a faster data rate (larger MCS), but at the same time, it increases the level of interference with surrounding APs. 
Additionally, most strategies were evaluated on relatively simple scenarios (with a few APs and a limited number of radio channels), often using a self-made network simulator.

\begin{table*}[t]
\caption{Comparison of the state-of-the-art data-driven strategies. The last column refers to the size of the scenarios involved in the performance evaluation of the strategy. For instance, 216/18 means the evaluation comprises 216 APs distributed over 18 radio channels.}
\label{tab:soa}
\centering
    \rowcolors{2}{gray!13}{white}
    \begin{tabular}{|l c c c c c l l|}
        \hline
        Proposed & Tuning of & \multicolumn{2}{c}{Degree of centralization} & Dynamic & {Traffic} & {Simulator} & {APs /} \\
        solution & {\texttt{TX\_PWR}} & Pull area & Push area & MCS & Up/Down & & {channels} \\
        \hline
         WCNC'15~\cite{dsc} &  & Associated STAs & {Associated STAs} &  & Up & Self-made & 100/3 \\
         WCNC'21~\cite{lsr} & \checkmark & {AP itself} & {AP itself} & \checkmark & Down & {ns-3} & 6/1\\
         Globecom'20~\cite{fsc} & & All APs & All APs &  & {Up/Down} & {ns-3} & 3/1\\
         ADHOC'19~\cite{sr_mab}  & \checkmark & {AP itself} & {AP itself} &  & Down & Self-made &  8/1\\
         JNCA'19~\cite{sr_mab_2}  & \checkmark & All APs & {AP itself} &  & Down & Self-made & 8/1 \\
         MSWiM'21~\cite{mab_mswim} & \checkmark & All APs & All APs &  & Down & {ns-3} & 10/1 \\
         \texttt{INSPIRE} & \checkmark & {Surrounding APs} & {Surrounding APs} & \checkmark & {Up/Down} & {ns-3} & {216/18} \\
        \hline
    \end{tabular}
\end{table*}

In this paper, we propose a fully distributed strategy to address the problem of the spatial reuse of radio channels in WLANs.  
The proposed strategy can be applied to any arrangement of WLANs and its novelties are mostly twofold. 
First, to the best of our knowledge, it is the first strategy making use of Gaussian Processes to explore promising WLAN configurations in the quest of discovering the optimal one. 
Gaussian processes are recognized tools to deal with the exploration vs. exploitation dilemma (see \cite{srinivas2009gaussian, chowdhury2017kernelized}) which is at the center of the spatial reuse problem.
Second, unlike the existing fully distributed strategies, \texttt{INSPIRE} allows each AP to account for its surroundings thanks to pull and push areas broader than a single node.
Through the use of a simple consensus method, APs of the WLANs achieve to behave altruistically selecting configurations for the ``greater good'' of the WLANs.
We also introduce realistic scenarios, inspired by real-life WLANs, with dynamic MCS and bidirectionnal traffic, to evaluate and compare the efficiency of all the considered strategies.

\section{Proposed solution} \label{sec:sol}

\subsection{WLANs under study}
Let $\mathcal{W}$ denote the set of concurrent WLANs under study, each of which being comprised of one or more APs.
We let $V$ be the set of APs in $\mathcal{W}$ that operate on the radio channel of interest. 
We denote by $N$ the number of APs, by $s_i$ the set of STAs associated with AP $i$ and by $S$ the total number of STAs in the considered radio channel of $\mathcal{W}$. 
Thus, we have: $S = \sum_{i = 1}^N |s_i|$. 
Finally, we use $\mathcal{N}_i$ to designate the set of APs that are within the communication range of AP $i$ (when every AP is under the default configuration of the \texttt{TX\_PWR} and \texttt{OBSS\_PD} parameters). Note that AP $i$ itself belongs to $\mathcal{N}_i$.
We refer to the APs in $\mathcal{N}_i$ as the surroundings of AP $i$. 

We make no assumptions on $\mathcal{W}$, including on the specific arrangement of its APs and STAs, other than the three detailed below. 

First, we assume that every AP $i$ is able to exchange control frames (possibly through its beacon frames) with its surrounding APs (i.e., the ones in $\mathcal{N}_i$). By the same token, we suppose that at least one AP $i$ has another AP in its communication range (\textit{i.e.}, $\exists i \in \llbracket1, N\rrbracket, \mathcal{N}_i \neq \{i\}$), otherwise the spatial reuse of the radio channel would already be at its apex.

Second, we assume that the $N$ APs have their \texttt{TX\_PWR} and \texttt{OBSS\_PD} parameters configurable (as it is the case since the introduction of the 802.11ax amendment). 
We use $x_i^t$ to denote the configuration of AP $i$ with regards to its two \texttt{TX\_PWR} and \texttt{OBSS\_PD} parameters at time $t$. 
Analogously, $x^t$ represents the configuration of the $N$ APs from $\mathcal{W}$ at time $t$.  
Thus, we have: $x_i^t \in C = \llbracket-82, -62\rrbracket \times \llbracket1, 21\rrbracket$ dBm and $x^t \in C^N$. 

Lastly, we assume that each AP in $\mathcal{W}$ can periodically run performance tests and obtain, in return, the mean throughput attained by each of its STAs over a short time interval $\Delta t$.
More formally, we use the vector $T^t \in \mathbb{R}^{+S}$ to denote the throughput attained by the $S$ STAs of $\mathcal{W}$ given the WLAN configuration $x^t$ at time $t$. Throughout this paper, we sometimes refer to $T^t$ as $T(x^t)$ to explicitly show the dependency between the STAs' throughputs and APs' configurations.

In this work, we seek to discover an adequate configuration $x^*$ of the $N$ APs composing $\mathcal{W}$ that improves the collective experience of the $S$ STAs through a better reuse of their radio channel. 
We address this problem as a reinforcement learning task in which, at regular time intervals $t$, the APs collect measurements $T^t$ associated to their current configuration $x^t$, and need to decide their next configuration $x^{t+1}$. 
The obstacles towards that objective are mostly threefold. 
(i) We need to define a meaningful objective function that APs will attempt to optimize collectively;
(ii) We are facing the well-known exploration vs. exploitation dilemma since the search for an adequate configuration of the WLANs should be as seamless as possible (without disrupting the STAs).
This leads us to cast the problem as a MAB problem where the arms are the WLANs' configurations. 
Following the MAB terminology, we refer to the objective function as the reward function; 
(iii) We are looking for a strategy that can be applied in a distributed way since it would be in general unrealistic to assume that (concurrent) APs have a fine knowledge beyond their surroundings.

\subsection{Reward function}
We need to define a reward function $R$ that appraises the ``goodness'' of a configuration $x$ with regards to the WLANs performance. 
Because multiple criteria may be considered in the definition of $R$, there is no universal definition. 
However, assessing the quality of a configuration $x$ can be derived from the STAs throughputs $T(x)$ obtained with APs configured with $x$. Among all easily computable reward functions, $R(x) = \prod_{T_i \in T(x)} T_i$ is called the proportional fairness (PF) and provides a convenient trade-off between fairness and cumulated throughput. However, PF is often criticized for its high variability, since $\frac{\partial R}{\partial T_i} = \prod_{T_j \in T(x), j \neq i} T_j$. 

To overcome this drawback, we consider the logarithm of PF.
This lowers its variability, which becomes: $\frac{\partial R}{\partial T_i} = \frac{1}{T_i}$ (note that $T_i$ is typically much larger than 1).  
This also emphasizes the contribution of STAs with low throughputs in the computation of $R$ and provides a pleasant closed-form to optimize. For an arbitrary set of APs $X$, we can define $R_X = \log \prod_{\substack{i \in X\\j \in s_i}} T_j(x)$. Then, our \textit{global} reward function $R$ is:
\begin{equation}
    \begin{split}
        R(x) = R_V(x) &= \log \prod_{\substack{i \in V\\j \in s_i}} T_j(x)\\
             &= \sum_{\substack{i \in V\\j \in s_i}} \log T_j(x)
    \end{split}
    \label{eq:reward}
\end{equation}

However, to compute Equation~\ref{eq:reward}, an AP must have a complete knowledge of the performance attained by the STAs of all APs or, at least, be able to communicate with all the APs in $\mathcal{W}$. 
This is in contradiction with our assumption that APs only have a partial knowledge of $\mathcal{W}$, limited to their surrounding APs. 
To design a reward function compatible with the distributed case, we proceed as follows. 
Each AP $i$ applies Equation~\ref{eq:reward} but restricted to the set of its associated STAs and obtains in return a ``selfish'' reward denoted by $R_{\{i\}}$. 
Previous work \cite{sr_mab} have showed that considering such selfish rewards may have a positive but limited impact on the WLANs performance. 
Therefore, we introduce a more altruistic reward, denoted by $R_i$ that accounts not only for the ``selfish'' reward of AP $i$ (i.e., $R_{\{i\}}$) but also for the rewards of the surrounding APs (i.e., the ones in $\mathcal{N}_i$).
The ``altruistic'' local reward of AP $i$ is computed as: 
\begin{equation}
    R_i(x) = \sum_{j \in \mathcal{N}_i} \frac{R_{\{j\}}(x)}{|\mathcal{N}_j|} 
    \label{eq:local_reward}
\end{equation}

Note that $R_i$ defined with Equation \ref{eq:local_reward} ensures that a configuration $x$ maximizing the local rewards is also a maximum for the global reward $R$ since we have: $\sum_{i = 1}^N R_i(x) = R(x)$.

\begin{proof}
    This is a straightforward property since
    \begin{equation}
        \sum_{i = 1}^N R_i(x) = \sum_{i=1}^N \sum_{j \in \mathcal{N}_i} \frac{R_{\{j\}}(x)}{|\mathcal{N}_j|}
    \end{equation}
    By noticing that $i \in \mathcal{N}_j \iff j \in \mathcal{N}_i$, we can permute the indices and use the definition of $R_{\{j\}}$:
    \begin{equation}
        \sum_{j = 1}^N \frac{R_{\{j\}}(x)}{|\mathcal{N}_j|} \sum_{i \in \mathcal{N}_j} 1 = \sum_{j = 1}^N \sum_{k \in s_j} \log T_k(x) = R(x)
        \label{eq:local_global}
    \end{equation}
\end{proof}

\subsection{Local reward maximization} \label{sec:reward_maxim}

For the sake of clarity and since all variables in this section are relative to an AP $i$, we often omit the subscript $i$ in the notations. 

Now that each AP $i$ has its own local reward function, we need a model of the knowledge of AP $i$ about $R_i$ in order to find its argmax. 
We represent the beliefs of AP $i$ about $R_i$ by defining a prior distribution on the reward functional space with a Gaussian Process (GP).

\textbf{Gaussian process.} In our case, a GP can be defined as a collection of random variables indexed by configurations of APs in $\mathcal{N}_i$: $\{Y_c; c \in C^{|\mathcal{N}_i|}\}$ such that every finite collection $\left(Y_{c_1}, \cdots, Y_{c_n}\right) \sim \mathcal{N}\left(\mu, \Sigma\right)$.
Without loss of generality, we assume the GP to have zero mean so that it is entirely determined by its covariance function $\Sigma : C^{|\mathcal{N}_k|} \times C^{|\mathcal{N}_k|} \rightarrow \mathbb{R}^+$. As shown by \cite{gp}, GPs can be used as priors on a function space.
We use $X_t$ to denote the $t \times 2|\mathcal{N}_k|$ features matrix gathering the tested configurations $\left(x^1, \cdots, x^t \right)^T$ and $Y_t$ to denote the $t \times 1$ label vector gathering the corresponding local reward values $\left(R_i(x^1), \cdots, R_i(x^t)\right)^T$. 
Given $X_t$ and $Y_t$,  we can infer the distribution of the reward value for an arbitrary configuration $x$, $R_i(x)$, from Bayes' theorem:
$R_i(x)|X_t,Y_t \sim \mathcal{N}\left(\mu(x), \sigma^2(x)\right)$ with $\mu(x)$  and $\sigma^2(x)$ defined in Equations \ref{eq:mu} and \ref{eq:sig}, respectively.
\begin{equation}
    \mu(x) = \Sigma(x, X_t)\Sigma(X_t, X_t)^{-1}Y_t
    \label{eq:mu}
\end{equation}
\begin{equation}
    \sigma^2(x) = \Sigma(x, x) - \Sigma(x, X_t)\Sigma(X_t, X_t)^{-1}\Sigma(X_t, x)
    \label{eq:sig}
\end{equation}

Since GPs can be used as a prior on a functional space, they are useful to solve regression problems as well as maximization tasks. In our case, the AP $i$ uses a GP to model $R_i$ and to assist the exploration of promising configurations of the APs in $\mathcal{N}_i$, maximizing $R_i$ in a Bayesian way.

Choosing the covariance function $\Sigma$ is a critical step when designing a GP as it determines some key features such as its isotropy and smoothness. Since the reward function, which quantifies the quality of spatial reuse in $\mathcal{N}_i$, is likely to exhibit threshold effects, we choose a covariance function that decreases rapidly as the distance between two considered configurations increases. 
Thus, the regularity constraint is not too restrictive on the modeled function. 
Because we have no incentive to prefer any particular direction over another, we let the covariance function $\Sigma(x, x')$ depend only on $||x - x'||$ to ensure the isotropy of the GP. 
This leads us to use a Matérn kernel \cite{matern} with parameter $\nu = \frac{3}{2}$, which is defined as 
\begin{equation}
    \Sigma(x, x') = s^2\left(1 + \frac{\sqrt{3}||x - x'||}{\rho}\right)e^{-\frac{\sqrt{3}||x - x'||}{\rho}}
    \label{eq:kernel}
\end{equation}

where $s^2$ and $\rho$ are two hyperparameters whose values are approximated by maximizing the likelihood of $Y_t$ (which is Gaussian) during the learning process. 

As discussed before, each AP $i$ faces the exploitation vs exploration dilemma in its attempt to find the optimal configuration. 
A common way in the MAB framework to appraise a given strategy $\pi$ is then to consider the cumulative regret $\Gamma(\pi)$. 
In our problem, $\Gamma(\pi)$ is expressed with Equation \ref{eq:cum_reg} for an episode of $D$ steps, since it is expressed as the cumulative sum of the differences between the best reward that the AP can get and $R_i(\pi(t))$, which is the actual reward obtained at time $t$ for the strategy $\pi$.

\begin{equation}
    \Gamma(\pi) = \sum_{t = 1}^D \max_{x \in C^{|\mathcal{N}_i|}} R_i(x) - R_i(\pi(t))
    \label{eq:cum_reg}
\end{equation}

Minimizing the cumulative regret with GP models is usually done by defining a strategy $\pi$ from the maximization of an acquisition function $A$: $\pi(t) = \argmax_{x \in C^{|\mathcal{N}_i|}} A_t(x)$. 
However, this assumes that our search space $C^{|\mathcal{N}_i|}$ is continuous. 
Since each AP $i$ deals with discrete configurations of APs in $\mathcal{N}_i$, we systematically round the recommendation of the GP to the nearest valid WLAN configuration. 
Many acquisition functions exist, such as Knowledge Gradient (KG) \cite{kg}, GP-UCB \cite{gpucb} or the Expected Improvement (EI) \cite{ei}. 
We choose EI over KG (whose computational cost can rapidly become prohibitive) and GP-UCB (which was found to be less efficient on our examples). 
The EI acquisition function is expressed as $A_t(x) = \mathbb{E}\left[(\mu_{t+1}(x) - \max_{1 \leq k \leq t} R_i(x_k))^+\right]$ given that $X_{t+1} = (X_t, x)$, $Y_{t+1} = (Y_t, R_i(x))$. Since $R_i(x) \sim \mathcal{N}(\mu(x), \sigma^2(x))$, we can derive a convenient closed-form for EI, as shown in Equation \ref{eq:ei}. 
\begin{equation}
    EI(x) = (\mu(x) - R_{i,t}^*) \Phi(Z) + \sigma(x)\phi(Z)
    \label{eq:ei}
\end{equation}

with $R_{i,t}^* = \max_{1 \leq k \leq t} R_i(x_k)$, $Z = \frac{\mu(x) - R_{i,t}^*}{\sigma(x)}$, $\Phi$ and $\phi$ being respectively the CDF and the PDF of a standard Gaussian distribution.

Then, the AP $i$ can try to maximize Equation \ref{eq:ei} by differentiating it and performing a gradient ascent.
By applying its strategy $\pi_i(t) = \argmax_{x \in C^{|\mathcal{N}_i|}} EI(x)$ and classical gradient ascent techniques on Equation \ref{eq:ei}, AP $i$ provides promising configurations for its surrounding APs in $\mathcal{N}_i$.

\subsection{Aggregation of local prescriptions}

In the previous sections, we have described how each AP $i$ computes its local reward and relies on its model $\mathcal{GP}_i$ to explore promising configurations for the APs in $\mathcal{N}_i$. 

\begin{figure}[t]
    \centering
    \includegraphics[width=0.87\linewidth]{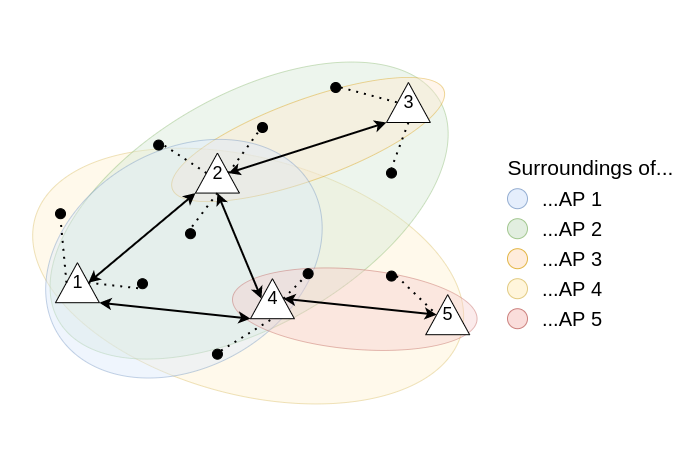}
    \caption{A WLAN represented by a graph with APs depicted as labelled triangles and STAs as black dots. An edge exists between two APs when they are in the communication range of each other. We use different colors to illustrate the surroundings of each AP in $\mathcal{F}$.}
    \label{fig:intersections}
\end{figure}

However, more coordination between APs is required. 
By construction, the collection $\mathcal{F} = \left(\mathcal{N}_k\right)_{1 \leq k \leq N}$ is a cover of the set of APs in $\mathcal{W}$ but not a partition.
In fact, if $\mathcal{F}$ had only null intersections (i.e., $\forall j,k, \mathcal{N}_j \cap \mathcal{N}_k = \emptyset$), then the spatial reuse of the radio channel would already be at its apex and there is no need for improvement. 
Figure \ref{fig:intersections} illustrates an example with 5 APs in which the collection $\mathcal{F} = (\mathcal{N}_1, \mathcal{N}_2, \mathcal{N}_3, \mathcal{N}_4, \mathcal{N}_5)$ exhibits multiple non-null intersections. 
As a result, most APs will receive \textit{a set} of \textit{different} prescriptions for the configuration of their \texttt{TX\_PWR} and \texttt{OBSS\_PD} parameters at their next iteration.
For instance, AP~1 will receive prescriptions from APs~2 and 4 in addition to its own prescription.
Since APs can only test one configuration at a time, one of those prescriptions must be chosen, or preferably, a consensus between them must be made.

In general, maximizing local rewards is very likely to lead to a sub-optimal situation since, for non-linear optimization problems, individual interests are often not aligned with the global interest (e.g., the famous Tragedy of the Commons \cite{tragedy_commons}). 
Without more information on the relation between the configuration of the APs and the measured throughputs of STAs, it seems difficult to provide an expression for the argmax of the global reward function $R$ given the argmax of the local reward functions $R_i$. However, recall that $\sum_{i = 1}^N R_i(x) = R(x)$. We can leverage this property to provide guarantees.

\begin{theorem}
    Let $\{R_i\}_{1 \leq i \leq N}$ be a set of Lipschitzian functions, $w \in [0, 1]^N, ||w||_1 = 1$ be a weight vector and $\{x^i_j\}_{i \in \mathcal{N}_j}$ be the prescriptions received by the AP $j$, with $x^i = \argmax_{x \in C^{|\mathcal{N}_i|}} R_i(x)$, then the weighted marginal median of the received sets of prescriptions is a minimax optimum $\Tilde{x}$ of $\sum_{i=1}^N R_i(\Tilde{x}_{\mathcal{N}_i})$:
    \begin{equation}
        \Tilde{x}_j = med\left(\{\left(x^i_j, w_i\right)\}_{i \in \mathcal{N}_j}\right)
        \label{eq:consensus}
    \end{equation}
    \label{thm:consensus}
\end{theorem}

\begin{proof}
    Let $x^i$ be the prescription of AP $i$, with $x^i_j$ the prescription of AP $i$ for AP $j$ and $w \in [0, 1]^n$ the weights of a convex combination ($||w||_1 = 1$). Let $R^* = \sum_{i = 1}^n R_i(x^i)$. If $x^i = \argmax_{x} R_i(x)$, then it follows that $\forall \Tilde{x} \in C^N, R^* - R(\Tilde{x}) \geq 0$. To find an optimal consensus $\Tilde{x}$, this difference must be minimized.
    
    If all the functions $R_i$ are Lipschitz-continuous with Lipschitz constants $L_i$, we have
    \begin{align}
        R^* - R(\Tilde{x}) &= \sum_{i = 1}^n R_i(x^i) - \sum_{i = 1}^n R_i(\Tilde{x}_{\mathcal{N}_i}) \nonumber\\
        &= \sum_{i = 1}^n R_i(x^i) - R_i(\Tilde{x}_{\mathcal{N}_i}) \nonumber\\
        &\leq \sum_{i = 1}^n L_i ||x^i - \Tilde{x}_{\mathcal{N}_i}||_1 \label{eq:inspire-thm_proof-lipschitz}\\
        &= \sum_{i = 1}^n L_i \sum_{j \in \mathcal{N}_i} \sum_{d = 1}^2 |x^i_{j,d} - \Tilde{x}_{j,d}| \label{eq:inspire-thm_proof-lipschitz_explicit}\\
        &= \Psi(\Tilde{x}) \nonumber
    \end{align}
    with~(\ref{eq:inspire-thm_proof-lipschitz}) following from each $R_i$ being Lipschitz continuous with Lipschitz constant $L_i$ and~(\ref{eq:inspire-thm_proof-lipschitz_explicit}) following from developing the norm.

    There is no lower upper bound of $R^* - R(x)$ than $\Psi(x)$, because we assume no information about $f$ other than its Lipschitz-continuity. Therefore, observe that if the minimizer of $\Psi(x)$ has a closed-form, it is the one of a minimax optimal. By rearranging the indices and splitting the absolute values we have
    \begin{align}
        \Psi(\Tilde{x}) &= \sum_{d = 1}^2 \sum_{j = 1}^n \sum_{i \in \mathcal{N}_j} L_i |x^i_{j,d} - \Tilde{x}_{j,d}| \nonumber\\
        &= \sum_{d = 1}^2 \sum_{j = 1}^n \left(\sum_{\substack{i \in \mathcal{N}_j\\x^i_{j,d} < \Tilde{x}_{j,d}}} L_i(\Tilde{x}_{j,d} - x^i_{j,d}) + \sum_{\substack{i \in \mathcal{N}_j\\x^i_{j,d} \geq \Tilde{x}_{j,d}}} L_i(x^i_{j,d} - \Tilde{x}_{j,d})\right). \nonumber\\
        &= \sum_{d = 1}^2 \sum_{j = 1}^n \psi_{j,d}(\Tilde{x}_{j,d})
    \end{align}

    Observe that $\Psi(\Tilde{x})$ is minimal if $\Tilde{x}_{j,d}$ is built to minimize $\psi_{j,d}(\Tilde{x}_{j,d})$. Therefore, let us focus on building $\Tilde{x}_{j,d}$ so that it minimizes
    \begin{equation} \label{eq:inspire-piecewise_linear}
        \psi_{j,d}(\Tilde{x}_{j,d}) = \sum_{\substack{i \in \mathcal{N}_j\\x^i_{j,d} < \Tilde{x}_{j,d}}} L_i(\Tilde{x}_{j,d} - x^i_{j,d}) + \sum_{\substack{i \in \mathcal{N}_j\\x^i_{j,d} \geq \Tilde{x}_{j,d}}} L_i(x^i_{j,d} - \Tilde{x}_{j,d}).
    \end{equation}
    
    It is immediate to see that~(\ref{eq:inspire-piecewise_linear}) is piecewise linear and convex since it is a sum of absolute values (\textit{i.e.} piecewise linear and convex functions). Furthermore, the set of nonlinearity points is necessarily $\mathcal{P}_{j,d} = \left\{x^i_{j,d}\right\}_{i \in \mathcal{N}_j}$. This directly implies that $\argmin_{x \in \mathbb{R}^+} \psi_{j,d}(x) \in \mathcal{P}_{j,d}$. Therefore, we need to find $k^* = \argmin_{k \in \mathcal{N}_j} \psi_{j,d}(x^k_{j,d})$. Without loss of generality, let us assume that $\mathcal{P}_{j,d}$ is sorted, so that $\forall k \in \mathcal{N}_j$,
    \begin{equation} \label{eq:inspire-piecewise_linear_rewritten}
        \psi_{j,d}(x^k_{j,d}) = \sum_{\substack{i \in \mathcal{N}_j\\i < k}} L_i(x^k_{j,d} - x^i_{j,d}) + \sum_{\substack{i \in \mathcal{N}_j\\i > k}} L_i(x^i_{j,d} - \Tilde{x}_{j,d}).
    \end{equation}

    Let $d_k = \psi_{j,d}(x^k_{j,d}) - \psi_{j,d}(x^{k+1}_{j,d})$. It is trivial to see that
    \begin{align} \label{eq:inspire-dk}
        d_k = \left(x^{k+1}_{j,d} - x^k_{j,d}\right) \left(\sum_{\substack{i \in \mathcal{N}_j\\i > k}} L_i - \sum_{\substack{i \in \mathcal{N}_j\\i < k+1}} L_i\right).
    \end{align}
    
    Since $\psi_{j,d}$ is convex, $k^*$ is necessarily the smallest $k$ for which $d_k$ is negative. Since $\left(x^{k+1}_{j,d} - x^k_{j,d}\right) \geq 0$ because $\mathcal{P}_{j,d}$ is sorted, $k^*$ is necessarily the smallest $k$ for which $\sum_{\substack{i \in \mathcal{N}_j\\i > k}} L_i - \sum_{\substack{i \in \mathcal{N}_j\\i < k+1}} L_i$ is negative. Therefore, we want $\Tilde{x}_{j,d} = \argmin_{x \in \mathbb{R}^+} \psi_{j,d}(x) = x^{k^*}_{j,d}$, which corresponds to the median of the values in $\mathcal{P}_{j, d}$, weighted by the Lipschitz constants $\left\{L_i\right\}_{i \in \mathcal{N}_j}$.

    Because each element in the vector $\Tilde{x} = \left(\Tilde{x}_{j,d}\right)_{d \in \left\{1, 2\right\}, j \in \left\{1, n\right\}}$ is built to be the weighted median of the prescriptions for the value of $x_{j,d}$, $\Tilde{x}$ is the weighted marginal median of the prescriptions. Eventually, since the weighted marginal median $\Tilde{x}$ of the prescriptions is indeed a minimum of $\Psi(x)$, it is a minimax optimum for the objective function $R$.
\end{proof}

Since, by definition of the local reward functions, $\sum_{i = 1}^N R_i(x) = R(x)$ (see Eq. \ref{eq:local_global}), we can apply Theorem \ref{thm:consensus} with, $\forall i \in \llbracket1, N\rrbracket, w_i = \frac{1}{N}$ and state that taking the marginal median of the prescriptions is a good way to reach high values of the global reward $R$.

\subsection{Algorithm and complexity}

\begin{algorithm}[h!]
    \caption{\texttt{INSPIRE} run at each AP $i$}
    \label{alg:solution}
    \hspace{-0.7cm}\textbf{Input}: subset $\mathcal{N}_i$ of APs
    \begin{algorithmic}[1]
        \STATE Initialize the Gaussian Process $\mathcal{GP}_i$
        \WHILE{\textbf{true}}
            \STATE Find a prescription $x^i = \argmax_{x \in C^{|\mathcal{N}_i|}} EI^t_i(x)$ by gradient ascent
            \STATE Broadcast $x^i$ to APs in $\mathcal{N}_i$
            \STATE Receive the prescriptions $x^j_i$ from AP $j, j \neq i, j \in \mathcal{N}_i$
            \STATE Compute the consensus $\Tilde{x}^{t+1}_i$ with Equation \ref{eq:consensus}
            \STATE Test $x^{t+1}_i$ for $\Delta t$ seconds and compute its selfish reward $R_{\{i\}}$ with Equation \ref{eq:reward} applied only to AP $i$
            \STATE Broadcast $R_{\{i\}}$, $|\mathcal{N}_i|$ and $\Tilde{x}^{t+1}_i$ to APs in $\mathcal{N}_i$
            \STATE Receive $R_{\{j\}}$, $|\mathcal{N}_j|$ and $\Tilde{x}^{t+1}_j$ from AP $j, j \neq i, j \in \mathcal{N}_i$
            \STATE Compute the local reward $R_i$ with Equation \ref{eq:local_reward} and the local configuration $\Tilde{x}^{t+1}_{\mathcal{N}_i}$
            \STATE Add the pattern $\left(\Tilde{x}^{t+1}_{\mathcal{N}_i}, R_i\right)$ to $\mathcal{GP}_k$
        \ENDWHILE
    \end{algorithmic}
\end{algorithm}

Algorithm \ref{alg:solution} summarizes the main steps of our proposed strategy \texttt{INSPIRE} run on each AP of the WLANs.

Contrary to what one might think, the most resource intensive operation in Algorithm \ref{alg:solution} is not the inversion of the $\Sigma(X_t, X_t)$ matrix. Since, at step $t$, we already know the Cholesky decomposition of $\Sigma(X_{t-1}, X_{t-1}) = LL^T$, it is very easy to compute the Cholesky decomposition of $\Sigma(X_{t}, X_{t})$. The most resource-intensive operation is the maximization of the acquisition function through gradient ascent. Since this requires computing many matrix-vector multiplications for at most $m$ steps, the computational complexity of Algorithm \ref{alg:solution} at time $t$ is $O\left(mt^2\right)$.

It is worth noting that the dimensionality of the problem (i.e. $\dim( C^{|\mathcal{N}_i|}) = |\mathcal{N}_k| \dim C$) does not appear in the expression of the asymptotic computational complexity of \texttt{INSPIRE}. 
This interesting property results from the use of a kernel function by GPs to compare WLANs configurations. 
This gives \texttt{INSPIRE} the ability to handle arbitrarily dense WLANs, or to optimize more parameters than just \texttt{TX\_PWR} and \texttt{OBSS\_PD}, without taking a hefty toll on its execution time. 
In fact, the real burden to the execution time of \texttt{INSPIRE} is $t$. 
This compels us to bound the size of $X_t$ and to find a balance between the amount of collected data on the WLANs' performance and configuration and a quick execution time. 
We keep the possibility of approximation methods to reduce the computational complexity of \texttt{INSPIRE} for future works. 
As for now, we recommend using windowing methods (such as a moving window) to bound the size of $X_t$ and so the computational complexity of \texttt{INSPIRE}.

\section{Performance Evaluation}

\subsection{Experimental settings}

To evaluate the ability of \texttt{INSPIRE} at improving the spatial reuse of a radio channel through the configuration of the \texttt{TX\_PWR} and \texttt{OBSS\_PD} parameters, we consider two distinct scenarios. 

\begin{figure*}
    \centering
        \begin{subfigure}{0.35\textwidth}
        \includegraphics[width=1\linewidth]{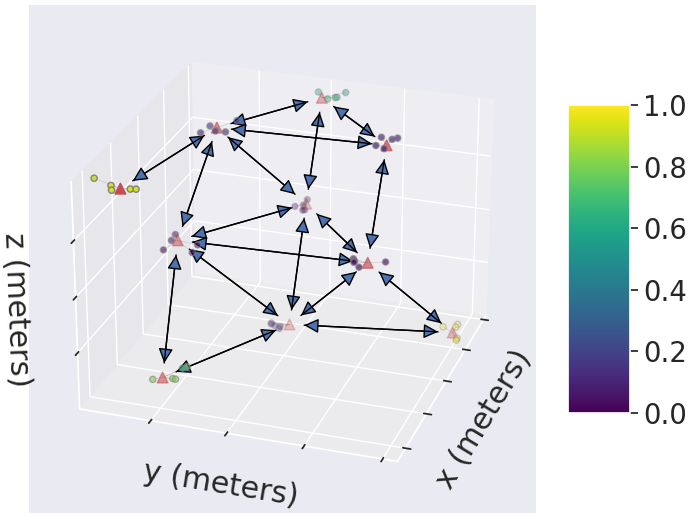}
        \caption{Topology \textbf{T1}.}
        \label{fig:topoT1}        
        \end{subfigure}
        \hspace{1.5cm}
        \begin{subfigure}{0.35\textwidth}
        \includegraphics[width=1\linewidth]{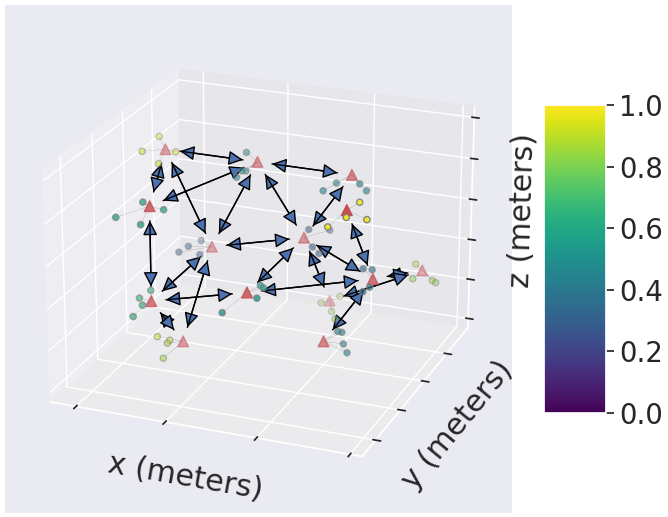}
        \caption{Topology \textbf{T2}.}
        \label{fig:topoT2}             
        \end{subfigure}
    \caption{The two considered topologies. APs are shown as red triangles, connected by a arrow if they lie in each other's communication range. Associated STAs are shown as dots colored according to their throughputs: warm, yellowish colors indicate that the STA has enough throughput most of the time while, on the contrary, cool, blueish colors indicate that the STA has mostly not enough throughput under the default configuration of 802.11: 20 dBm for \texttt{TX\_PWR} and -82 dBm for \texttt{OBSS\_PD}.} 
    \label{fig:topos}
\end{figure*}

The first scenario is inspired by the WLAN deployment of Cisco in their offices in San Francisco. 
In~\cite{cisco_topo}, Cisco provides the location of 60 APs that together deliver wireless connectivity to their employees on a floor.  
To account for the WLANs' activity from other floors, we consider a three-floor building and we replicate on each floor the same arrangement of APs as in Cisco's offices. 
This leads us to a total number of 180 APs spanned over three floors. 
Assuming 18 independent radio channels, we run a radio channel allocation algorithm to determine the radio channel used by each  AP. 
For our first scenario, we consider the subgraph resulting from the channel allocation with the highest density. 
We use \textbf{T1} to refer to this topology (i.e., arrangements of APs and STAs), which is illustrated in Figure~\ref{fig:topoT1}.  
\textbf{T1} exhibits a total of 10 APs and we associate a number of 5 STAs per AP. 

The second scenario addresses the case of many single-AP WLANs deployed and operated independently in a relatively limited area. 
This is typically the case in housing units where each apartment is equipped with its own AP so that the APs are often only a few meters away from a number of others. 
More specifically, we consider a nine-story building with 216 apartments of 25 m² each. 
We randomly position an AP within each apartment as well as 4 STAs per AP. 
Then, similarly to the first scenario, we apply a radio channel allocation algorithm given a total of 18 radio channels, to obtain the topology of interest denoted by \textbf{T2}. 
Note that \textbf{T2} consists of 14 APs and 56 STAs. 
Figure~\ref{fig:topoT2} depicts the topology \textbf{T2}. 


For each scenario, we consider heavily loaded conditions. 
APs attempt to transmit frames to each of their associated STAs at a rate of 50 Mbps while the latter attempt to send their frames to the AP at a lower rate of 3.33 Mbps. 
These assumptions are in line with the downstream traffic largely exceeding the upstream traffic in WLANs. 
Given the speed of wireless links in 802.11ax, the buffers of the APs will always be full of frames waiting to be sent. 
More generally, considering APs in saturation represents undoubtedly the most difficult case when dealing with the spatial reuse of a radio channel. 
Therefore, if \texttt{INSPIRE} manages to significantly improve the WLANs' performance under these circumstances, then it can only do better under normal conditions.

To better appraise the quality of \texttt{INSPIRE}, we also consider a control strategy as well as several state-of-the-art solutions, which were discussed in Section~\ref{sec:soa} and briefly summarized here: 
\begin{itemize}
    \item \texttt{DEFAULT}: Every AP keeps its default configuration for the \texttt{TX\_PWR} and \texttt{OBSS\_PD} parameters (i.e., $(-82, 20)$ dBm);
    \item \texttt{WCNC'15}: Each AP implements a simple distributed algorithm to dynamically update its \texttt{OBSS\_PD} parameter \cite{dsc};
    \item \texttt{JNCA'19}: Each AP solves a MAB problem using Thompson sampling to dynamically update their \texttt{TX\_PWR} and \texttt{OBSS\_PD} parameters \cite{sr_mab_2};
    \item \texttt{MSWiM'21}: Similar to \texttt{JNCA'19}, except that the sampling of new configurations is performed through a multivariate Gaussian mixture, and that the solution is centralized \cite{mab_mswim}.
\end{itemize}

We implemented \texttt{INSPIRE} (based on the open-source Gaussian process library LibGP \cite{libgp}) as well as the four strategies described above in the open-source network simulator ns-3 \cite{ns3}. 
ns-3 is a well-established realistic discrete-event simulator that implements most of the network protocols involved in the WLANs communication from the Physical up to the Application layer. 
We report in Table~\ref{tab:ns3_parameters} the simulation parameters used in the rest of this section. 
Unlike previous works (e.g., \cite{mab_mswim, dsc, fsc, sr_mab, sr_mab_2})) with the exception of \cite{lsr}, our simulations incorporate the mechanism of rate adaptation that let APs and STAs dynamically vary the speed of their wireless links (through the use of different Modulation Coding Scheme (MCS)) in response to the quality of the received signal. 
This is particularly important for the sake of our study since changing the value of \texttt{TX\_PWR} necessarily affects the quality of the received signal and thus the MCS. 
Since our simulated WLANs take place in buildings, we choose an appropriate path loss by combining the models \texttt{ItuR1238} and \texttt{InternalWallsLoss}, both implemented by ns-3. 
With these propagation models, the signal is decreased by an additional attenuation coefficient each time it goes through a floor or a wall. The attenuation coefficients are respectively -4 dBm (which is the default value in \texttt{ItuR1238}) and -8 dBm.

\begin{table}[t]
\caption{ns-3 parameters.}
\label{tab:ns3_parameters}
\centering
    \rowcolors{2}{gray!13}{white}
    \begin{tabular}{|l p{4cm}|}
        \hline
       \textbf{Parameter} & \textbf{Value}\\
       \hline
       ns-3 version & 3.31\\
       Number of repetitions & 22\\
       Simulation duration & 30 s\\
       Duration of an iteration ($\Delta t$) & 75 ms\\
       Packet size & 1,464 bytes\\
       Downlink traffic & 50.0 Mbps\\
       Uplink traffic & 3.33 Mbps\\
       Channel size & 20 MHz\\
       Frequency band & 5 GHz\\
       A-MDPU Aggregation & 4\\
       Path loss & \texttt{HybridBuildings} (\texttt{ItuR1238} + \texttt{InternalWallsLoss})\\
       Wi-Fi Manager & \texttt{IdealWifiManager}\\
       \hline
    \end{tabular}
\end{table}

We instrumented ns-3 to collect and compute a number of performance metrics. 
At the end of each iteration, the quality of the spatial reuse is assessed with Equation \ref{eq:reward}, although distributed strategies may internally use the local reward defined in Equation~\ref{eq:local_reward}. 
Then, we compute the classical metric to analyze the efficiency of a strategy at dealing with a MAB problem: (i) The cumulative regret (with Equation \ref{eq:cum_reg} using a normalized version of the global reward in Equation~\ref{eq:reward}). 
We also collect the following performance metrics to reflect the effect of each strategy on the behavior of the WLANs and of their STAs: (ii) The number of starving STAs, which we define as STAs experiencing a very low throughput (namely, less than 10\% of their attainable throughput) and (iii) The cumulated throughput, which simply sums all STAs' throughput.

Each simulation lasts 30 seconds and we replicated them independently 22 times to obtain and visualize their first, second, and third quartiles. 
When the quartiles of a performance metric vary too much within a single simulation, we apply an exponential moving average (with $\alpha = 0.04$) to extract the underlying trends of the quartiles sequences. 
The metrics are collected throughout the whole duration of the simulation. 
At the end of each iteration, we compute all the performance metrics and then we refer to the current strategy to decide what will be the next configuration of the WLANs.
Since an iteration lasts $\Delta t$ = 75 ms and a simulation lasts 30 seconds, the quality of each solution is assessed over 400 iterations.

\subsection{Numerical results} \label{sec:num_res}

\begin{figure*}[t]
    \centering
    \begin{subfigure}{0.33\textwidth}
        \centering
        \includegraphics[width=\linewidth]{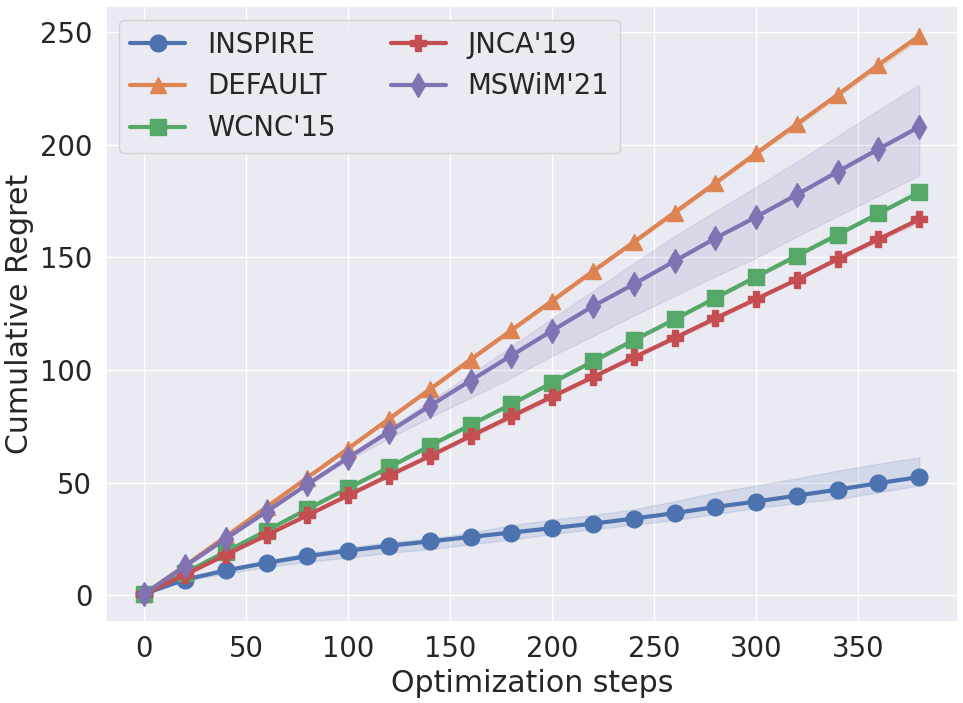}
        \caption{Cumulative Regret} 
        \label{fig:t1_cumreg}
    \end{subfigure}
    \begin{subfigure}{0.33\textwidth}
        \centering
        \includegraphics[width=\textwidth]{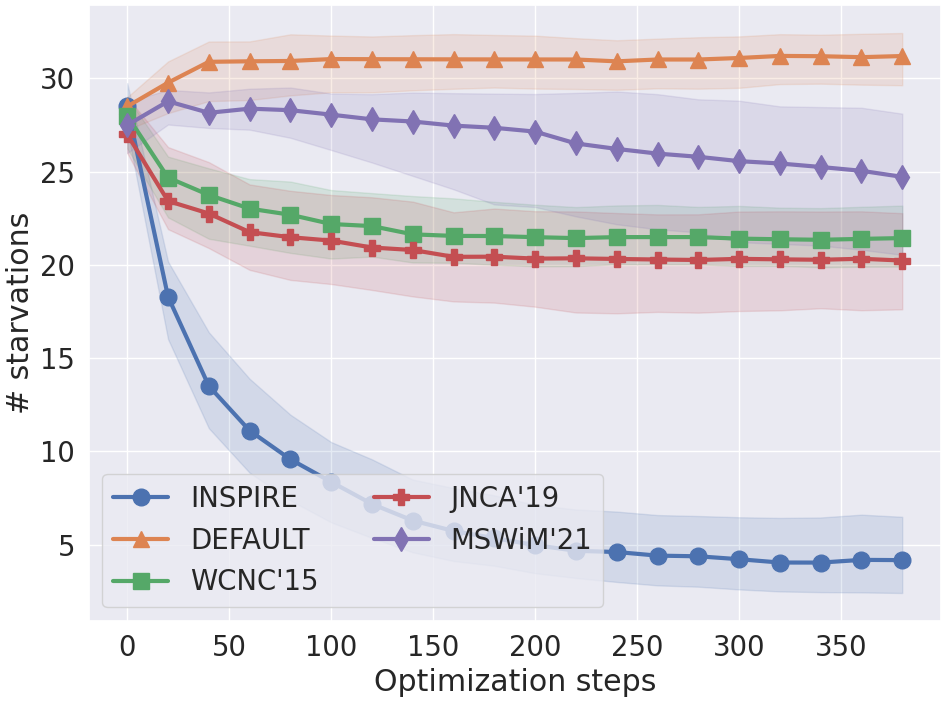}
        \caption{Starvations}
        \label{fig:t1_starv}
    \end{subfigure}
    \begin{subfigure}{0.33\textwidth}
        \centering
        \includegraphics[width=\textwidth]{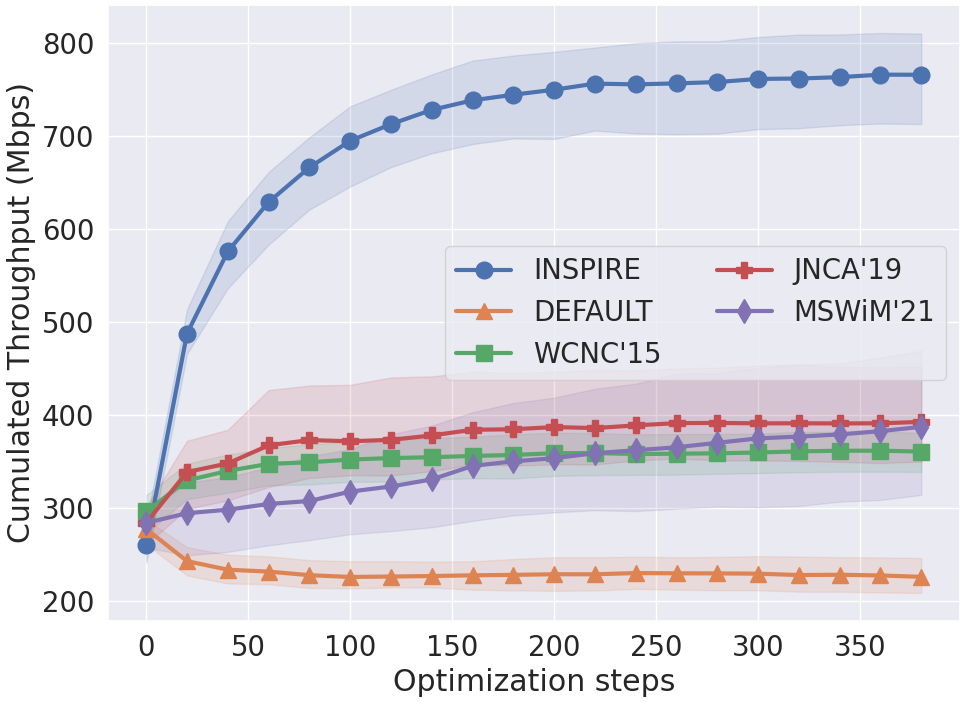}
        \caption{Cumulated Throughput}
        \label{fig:t1_cumthrough}
    \end{subfigure}
    \caption{Performance metrics delivered on topology \textbf{T1} by each strategy.}
    \label{fig:t1_res}
\end{figure*}

Figure~\ref{fig:t1_res} illustrates the performance metrics delivered by the ns-3 simulator for each strategy in the case of topology \textbf{T1}. 
The cumulative regret, represented in Figure \ref{fig:t1_cumreg}, indicates which strategy has performed the best at any time of the simulation. 
\texttt{INSPIRE} is found to be the most efficient strategy, reducing the cumulative regret by 70\% more than \texttt{DEFAULT} and by over 50\% than \texttt{WCNC'15}, which happens to be the most efficient state-of-the-art strategy.
We now look at the other performance metrics to better understand how much \texttt{INSPIRE} is able to improve the behavior of the WLAN and of its STAs. 
Taking \texttt{DEFAULT} as baseline, Figure \ref{fig:t1_starv} shows that \texttt{INSPIRE} reduces the number of STAs in starvation by 80\% while Figure \ref{fig:t1_cumthrough} demonstrate that our proposed solution manages to increase the cumulated throughput (+400\%).

\begin{figure*}[t]
    \centering
    \begin{subfigure}{0.33\textwidth}
        \centering
        \includegraphics[width=\linewidth]{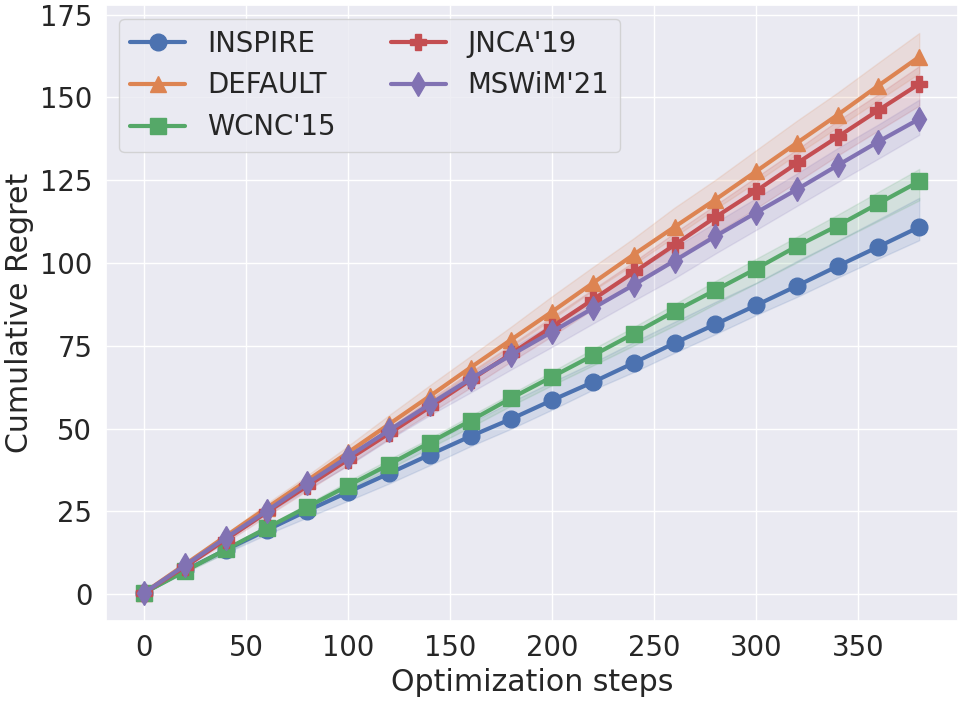}
        \caption{Cumulative Regret} 
        \label{fig:t2_cumreg}
    \end{subfigure}
    \begin{subfigure}{0.33\textwidth}
        \centering
        \includegraphics[width=\textwidth]{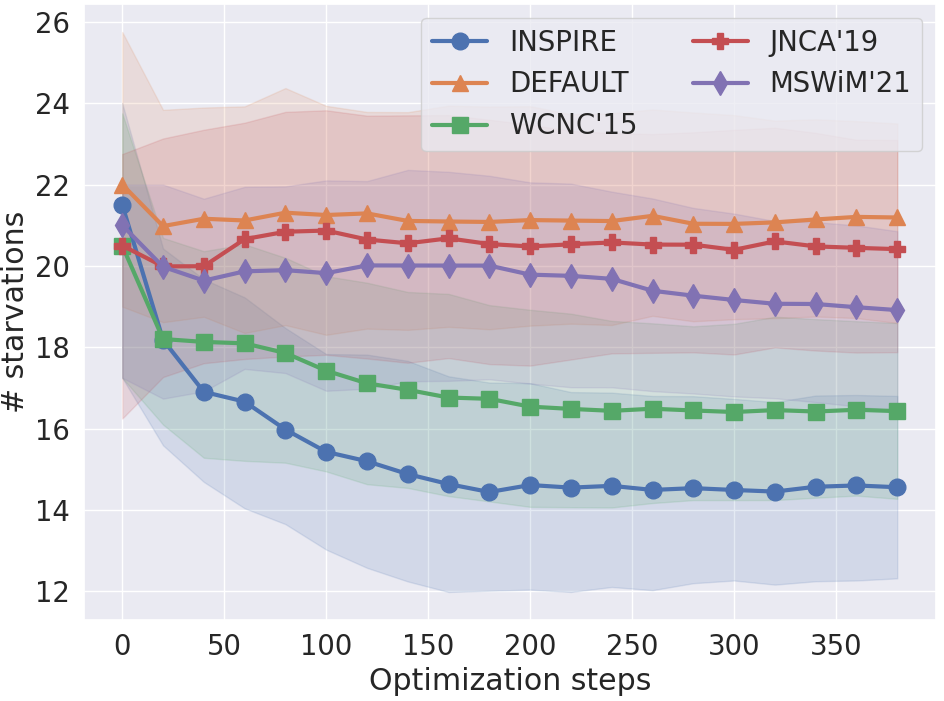}
        \caption{Starvations}
        \label{fig:t2_starv}
    \end{subfigure}
    \begin{subfigure}{0.33\textwidth}
        \centering
        \includegraphics[width=\textwidth]{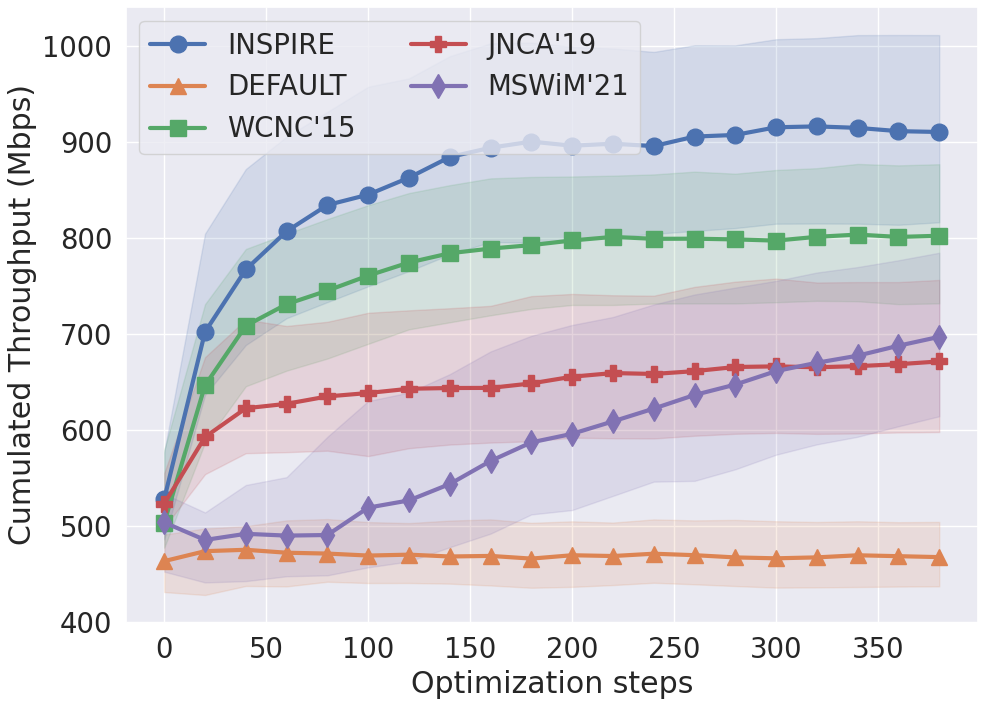}
        \caption{Cumulated Throughput}
        \label{fig:t2_cumthrough}
    \end{subfigure}
    \caption{Performance metrics delivered on topology \textbf{T2} by each strategy.}
    \label{fig:t2_res}
\end{figure*}

We now turn to the case of topology \textbf{T2}. 
First, we observe in Figure \ref{fig:t2_cumreg} that among the four considered strategies, \texttt{INSPIRE} is the one that manages to decrease the most the cumulative regret with a decline of about 36\% compared to the \texttt{DEFAULT} configuration at the end of the simulation.
The proposed solution also outperforms \texttt{MSWiM'21}, which is found to be the best state-of-the-art strategy on this topology, by a margin of 14\%.
Looking at the performance of WLANs and of their STAs,  Figure~\ref{fig:t2_starv} shows that \texttt{INSPIRE} is able to limit the number of STAs starving from throughput by a degree of 36\% when compared to the \texttt{DEFAULT} configuration. 
Similarly, the cumulated throughput of STAs have their value increased by 28\% and nearly doubled with \texttt{INSPIRE} (see Figure \ref{fig:t2_cumthrough}). 

Overall, through the study of topologies \textbf{T1} and \textbf{T2}, \texttt{INSPIRE} demonstrates its superiority over the other state-of-the-art strategies. The significant improvements brought by our proposed solution on all performance metrics are permanently obtained after 100 iterations only (7.5 seconds). 
In other words, in less than 10 seconds, \texttt{INSPIRE} manages to significantly improve the behavior of the WLANs and of the associated STAs thanks to a better spatial reuse of the radio channel.
This efficiency in searching and finding an adequate configuration of the \texttt{TX\_PWR} and \texttt{OBSS\_PD} parameters at each AP of the WLANs mostly results from the distributed, altruistic use of Gaussian processes which we further discuss in the next section. 

\section{Conclusions} \label{sec:conc}

In this work, we have presented \texttt{INSPIRE}, a reinforcement learning method to improve the spatial reuse of radio channels in WLANs by configuring two parameters of APs: the transmission power (\texttt{TX\_PWR}) and the sensitivity threshold (\texttt{OBSS\_PD}), that can be dynamically configured with the latest Wi-Fi amendments. 
To address the difficult problem of sharing efficiently and fairly the resource of a radio channel, \texttt{INSPIRE} works as a distributed solution where each AP solves a local Multi-Armed Bandit problem with the help of information and actions limited to its surrounding APs (\textit{i.e}. within its communication range). 
The development of the solution includes (i) an intuitive quantification (based on STAs throughputs) of the ``goodness'' of a configuration of \texttt{TX\_PWR} and \texttt{OBSS\_PD} for concurrent APs in WLANs, both at local and global scales, (ii) the use of an acquisition function and Gaussian processes to find local configurations that maximize approximations of local reward functions, and (iii) an altruistic behavior facilitated by prescriptions to surrounding APs along with a consensus method which aggregates the prescriptions of surrounding APs for the ``greater good'' of the WLANs.

\texttt{INSPIRE} has been evaluated and compared with other state-of-the-art strategies addressing the same problem, using the open-source network simulator ns-3 that implements all the layers of the network stack. 
The different strategies were compared on two examples inspired by real-life deployments of dense WLANs in both professional and domestic environments. 
\texttt{INSPIRE} was found to outperform other state-of-the-art strategies by significantly reducing the number of STAs in starvation and increasing the cumulated throughput of the WLANs in only a few seconds.

As future work, we plan to assess the quality of \texttt{INSPIRE} for a specific class of WLANs where STAs are mobile (e.g., customers in a shopping mall). 
Another natural follow-up would be to experiment \texttt{INSPIRE} with real material on a testbed.

\begin{acks}
    This work was supported by the LABEX MILYON (ANR-10-LABX-0070) of Universit\'e de Lyon, within the program ``Investissements d'Avenir'' (ANR-11-IDEX- 0007) operated by the French National Research Agency (ANR).
\end{acks}

\bibliographystyle{ACM-Reference-Format}
\bibliography{main}


\begin{thebibliography}{24}


\ifx \showCODEN    \undefined \def \showCODEN     #1{\unskip}     \fi
\ifx \showDOI      \undefined \def \showDOI       #1{#1}\fi
\ifx \showISBNx    \undefined \def \showISBNx     #1{\unskip}     \fi
\ifx \showISBNxiii \undefined \def \showISBNxiii  #1{\unskip}     \fi
\ifx \showISSN     \undefined \def \showISSN      #1{\unskip}     \fi
\ifx \showLCCN     \undefined \def \showLCCN      #1{\unskip}     \fi
\ifx \shownote     \undefined \def \shownote      #1{#1}          \fi
\ifx \showarticletitle \undefined \def \showarticletitle #1{#1}   \fi
\ifx \showURL      \undefined \def \showURL       {\relax}        \fi
\providecommand\bibfield[2]{#2}
\providecommand\bibinfo[2]{#2}
\providecommand\natexlab[1]{#1}
\providecommand\showeprint[2][]{arXiv:#2}

\bibitem[\protect\citeauthoryear{??}{802}{2021a}]%
        {802.11-2020}
 \bibinfo{year}{2021}\natexlab{a}.
\newblock \showarticletitle{{IEEE} Standard for Information
  Technology--Telecommunications and Information Exchange between Systems -
  Local and Metropolitan Area Networks--Specific Requirements - Part 11:
  Wireless LAN Medium Access Control (MAC) and Physical Layer (PHY)
  Specifications}.
\newblock \bibinfo{journal}{\emph{{IEEE} Std 802.11-2020 (Revision of IEEE Std
  802.11-2016)}} (\bibinfo{year}{2021}).
\newblock


\bibitem[\protect\citeauthoryear{??}{802}{2021b}]%
        {802.11-2021}
 \bibinfo{year}{2021}\natexlab{b}.
\newblock \showarticletitle{IEEE Standard for Information
  Technology--Telecommunications and Information Exchange between Systems Local
  and Metropolitan Area Networks--Specific Requirements Part 11: Wireless LAN
  Medium Access Control (MAC) and Physical Layer (PHY) Specifications Amendment
  1: Enhancements for High-Efficiency WLAN}.
\newblock \bibinfo{journal}{\emph{IEEE Std 802.11ax-2021 (Amendment to IEEE Std
  802.11-2020)}} (\bibinfo{year}{2021}), \bibinfo{pages}{1--767}.
\newblock
\urldef\tempurl%
\url{https://doi.org/10.1109/IEEESTD.2021.9442429}
\showDOI{\tempurl}


\bibitem[\protect\citeauthoryear{Afaqui, Garcia-Villegas, Lopez-Aguilera,
  Smith, and Camps}{Afaqui et~al\mbox{.}}{2015}]%
        {dsc}
\bibfield{author}{\bibinfo{person}{M.~Shahwaiz Afaqui}, \bibinfo{person}{Eduard
  Garcia-Villegas}, \bibinfo{person}{Elena Lopez-Aguilera},
  \bibinfo{person}{Graham Smith}, {and} \bibinfo{person}{Daniel Camps}.}
  \bibinfo{year}{2015}\natexlab{}.
\newblock \showarticletitle{Evaluation of dynamic sensitivity control algorithm
  for IEEE 802.11ax}. In \bibinfo{booktitle}{\emph{IEEE Wireless Communications
  and Networking Conference (WCNC'15).}} \bibinfo{pages}{1060--1065}.
\newblock
\urldef\tempurl%
\url{https://doi.org/10.1109/WCNC.2015.7127616}
\showDOI{\tempurl}


\bibitem[\protect\citeauthoryear{Ak and Canberk}{Ak and Canberk}{2020}]%
        {fsc}
\bibfield{author}{\bibinfo{person}{Elif Ak} {and} \bibinfo{person}{Berk
  Canberk}.} \bibinfo{year}{2020}\natexlab{}.
\newblock \showarticletitle{FSC: Two-Scale AI-Driven Fair Sensitivity Control
  for 802.11ax Networks}. In \bibinfo{booktitle}{\emph{IEEE Global
  Communications Conference (GLOBECOM'20).}} \bibinfo{pages}{1--6}.
\newblock
\urldef\tempurl%
\url{https://doi.org/10.1109/GLOBECOM42002.2020.9322153}
\showDOI{\tempurl}


\bibitem[\protect\citeauthoryear{Bardou, Begin, and Busson}{Bardou
  et~al\mbox{.}}{2021}]%
        {mab_mswim}
\bibfield{author}{\bibinfo{person}{Anthony Bardou}, \bibinfo{person}{Thomas
  Begin}, {and} \bibinfo{person}{Anthony Busson}.}
  \bibinfo{year}{2021}\natexlab{}.
\newblock \showarticletitle{Improving the Spatial Reuse in IEEE 802.11ax WLANs:
  A Multi-Armed Bandit Approach}. In \bibinfo{booktitle}{\emph{ACM
  International Conference on Modeling, Analysis and Simulation of Wireless and
  Mobile Systems (MSWiM'21)}}.
\newblock


\bibitem[\protect\citeauthoryear{Blum}{Blum}{2016}]%
        {libgp}
\bibfield{author}{\bibinfo{person}{Manuel Blum}.}
  \bibinfo{year}{2016}\natexlab{}.
\newblock \bibinfo{title}{LibGP}.
\newblock \bibinfo{howpublished}{GitHub, https://github.com/mblum/libgp}.
\newblock


\bibitem[\protect\citeauthoryear{Chowdhury and Gopalan}{Chowdhury and
  Gopalan}{2017}]%
        {chowdhury2017kernelized}
\bibfield{author}{\bibinfo{person}{Sayak~Ray Chowdhury} {and}
  \bibinfo{person}{Aditya Gopalan}.} \bibinfo{year}{2017}\natexlab{}.
\newblock \showarticletitle{On kernelized multi-armed bandits}. In
  \bibinfo{booktitle}{\emph{International Conference on Machine Learning}}.
  PMLR, \bibinfo{pages}{844--853}.
\newblock


\bibitem[\protect\citeauthoryear{Cisco}{Cisco}{2020}]%
        {cisco_topo}
\bibfield{author}{\bibinfo{person}{Cisco}.} \bibinfo{year}{2020}\natexlab{}.
\newblock \bibinfo{title}{High Density Wi-Fi Deployments}.
\newblock
  \bibinfo{howpublished}{\url{https://documentation.meraki.com/Architectures_and_Best_Practices/Cisco_Meraki_Best_Practice_Design/Best_Practice_Design_-_MR_Wireless/High_Density_Wi-Fi_Deployments}}.
\newblock
\newblock
\shownote{Accessed: 2022-01-21.}


\bibitem[\protect\citeauthoryear{Cisco}{Cisco}{2019}]%
        {cisco2019wp}
\bibfield{author}{\bibinfo{person}{VNI Cisco}.}
  \bibinfo{year}{2019}\natexlab{}.
\newblock \showarticletitle{Cisco visual networking index: Forecast and trends,
  2017--2022 white paper}.
\newblock \bibinfo{journal}{\emph{Cisco Internet Report}}  \bibinfo{volume}{17}
  (\bibinfo{year}{2019}), \bibinfo{pages}{13}.
\newblock


\bibitem[\protect\citeauthoryear{Genton}{Genton}{2002}]%
        {matern}
\bibfield{author}{\bibinfo{person}{Marc~G. Genton}.}
  \bibinfo{year}{2002}\natexlab{}.
\newblock \showarticletitle{Classes of kernels for machine learning: a
  statistics perspective}.
\newblock \bibinfo{journal}{\emph{Journal of Machine Learning Research}}
  \bibinfo{volume}{2} (\bibinfo{date}{3} \bibinfo{year}{2002}).
\newblock


\bibitem[\protect\citeauthoryear{Gupta and Miescke}{Gupta and Miescke}{1996}]%
        {kg}
\bibfield{author}{\bibinfo{person}{Shanti~S. Gupta} {and}
  \bibinfo{person}{Klaus~J. Miescke}.} \bibinfo{year}{1996}\natexlab{}.
\newblock \showarticletitle{Bayesian look ahead one-stage sampling allocations
  for selection of the best population}.
\newblock \bibinfo{journal}{\emph{Journal of Statistical Planning and
  Inference}} \bibinfo{volume}{54}, \bibinfo{number}{2} (\bibinfo{year}{1996}),
  \bibinfo{pages}{229--244}.
\newblock
\showISSN{0378-3758}
\urldef\tempurl%
\url{https://doi.org/10.1016/0378-3758(95)00169-7}
\showDOI{\tempurl}


\bibitem[\protect\citeauthoryear{Hardin}{Hardin}{2009}]%
        {tragedy_commons}
\bibfield{author}{\bibinfo{person}{Garrett Hardin}.}
  \bibinfo{year}{2009}\natexlab{}.
\newblock \showarticletitle{The Tragedy of the Commons}.
\newblock \bibinfo{journal}{\emph{Journal of Natural Resources Policy
  Research}} \bibinfo{volume}{1}, \bibinfo{number}{3} (\bibinfo{year}{2009}),
  \bibinfo{pages}{243--253}.
\newblock
\urldef\tempurl%
\url{https://doi.org/10.1080/19390450903037302}
\showDOI{\tempurl}


\bibitem[\protect\citeauthoryear{Jones, Schonlau, and Welch}{Jones
  et~al\mbox{.}}{1998}]%
        {ei}
\bibfield{author}{\bibinfo{person}{Donald~R. Jones}, \bibinfo{person}{Matthias
  Schonlau}, {and} \bibinfo{person}{William~J. Welch}.}
  \bibinfo{year}{1998}\natexlab{}.
\newblock \showarticletitle{Efficient Global Optimization of Expensive
  Black-Box Functions}.
\newblock \bibinfo{journal}{\emph{Journal of Global Optimization}}
  \bibinfo{volume}{13} (\bibinfo{year}{1998}), \bibinfo{pages}{455–492}.
\newblock
\urldef\tempurl%
\url{https://doi.org/10.1023/A:1008306431147}
\showDOI{\tempurl}


\bibitem[\protect\citeauthoryear{Kim, Yu, and Choi}{Kim et~al\mbox{.}}{2004}]%
        {Kim2004}
\bibfield{author}{\bibinfo{person}{Youngsoo Kim}, \bibinfo{person}{Jeonggyun
  Yu}, {and} \bibinfo{person}{Sunghyun Choi}.} \bibinfo{year}{2004}\natexlab{}.
\newblock \showarticletitle{{SP-TPC}: a self-protective energy efficient
  communication strategy for {IEEE} 802.11 {WLANs}}. In
  \bibinfo{booktitle}{\emph{IEEE Vehicular Technology Conference (VTC'04)}}.
\newblock


\bibitem[\protect\citeauthoryear{Lee, Kim, and Bahk}{Lee et~al\mbox{.}}{2021}]%
        {lsr}
\bibfield{author}{\bibinfo{person}{Hyunjoong Lee}, \bibinfo{person}{Hyung-Sin
  Kim}, {and} \bibinfo{person}{Saewoong Bahk}.}
  \bibinfo{year}{2021}\natexlab{}.
\newblock \showarticletitle{LSR: Link-aware Spatial Reuse in IEEE 802.11ax
  WLANs}. In \bibinfo{booktitle}{\emph{IEEE Wireless Communications and
  Networking Conference (WCNC'21)}}. \bibinfo{pages}{1--6}.
\newblock
\urldef\tempurl%
\url{https://doi.org/10.1109/WCNC49053.2021.9417353}
\showDOI{\tempurl}


\bibitem[\protect\citeauthoryear{ns-3.31.}{ns-3.31.}{2020}]%
        {ns3}
ns-3.31. \bibinfo{year}{2020}\natexlab{}.
\newblock \bibinfo{title}{{T}he {N}etwork {S}imulator ns-3}.
\newblock \bibinfo{howpublished}{\url{https://www.nsnam.org/}}.
\newblock
\newblock
\shownote{Accessed: 2022-01-21.}


\bibitem[\protect\citeauthoryear{Qiu, Chu, Leung, and Kee-Yin~Ng}{Qiu
  et~al\mbox{.}}{2020}]%
        {Qiu2020}
\bibfield{author}{\bibinfo{person}{Shuwei Qiu}, \bibinfo{person}{Xiaowen Chu},
  \bibinfo{person}{Yiu-Wing Leung}, {and} \bibinfo{person}{Joseph Kee-Yin~Ng}.}
  \bibinfo{year}{2020}\natexlab{}.
\newblock \showarticletitle{Joint Access Point Placement and
  Power-Channel-Resource-Unit Assignment for 802.11ax-Based Dense WiFi with QoS
  Requirements}. In \bibinfo{booktitle}{\emph{IEEE Conference on Computer
  Communications (INFOCOM'20).}} \bibinfo{pages}{2569--2578}.
\newblock
\urldef\tempurl%
\url{https://doi.org/10.1109/INFOCOM41043.2020.9155490}
\showDOI{\tempurl}


\bibitem[\protect\citeauthoryear{Srinivas, Krause, Kakade, and Seeger}{Srinivas
  et~al\mbox{.}}{2009}]%
        {srinivas2009gaussian}
\bibfield{author}{\bibinfo{person}{Niranjan Srinivas}, \bibinfo{person}{Andreas
  Krause}, \bibinfo{person}{Sham~M. Kakade}, {and} \bibinfo{person}{Matthias
  Seeger}.} \bibinfo{year}{2009}\natexlab{}.
\newblock \showarticletitle{Gaussian process optimization in the bandit
  setting: No regret and experimental design}.
\newblock \bibinfo{journal}{\emph{arXiv preprint arXiv:0912.3995}}
  (\bibinfo{year}{2009}).
\newblock


\bibitem[\protect\citeauthoryear{Srinivas, Krause, Kakade, and Seeger}{Srinivas
  et~al\mbox{.}}{2012}]%
        {gpucb}
\bibfield{author}{\bibinfo{person}{Niranjan Srinivas}, \bibinfo{person}{Andreas
  Krause}, \bibinfo{person}{Sham~M. Kakade}, {and} \bibinfo{person}{Matthias~W.
  Seeger}.} \bibinfo{year}{2012}\natexlab{}.
\newblock \showarticletitle{Information-Theoretic Regret Bounds for Gaussian
  Process Optimization in the Bandit Setting}.
\newblock \bibinfo{journal}{\emph{IEEE Transactions on Information Theory}}
  \bibinfo{volume}{58}, \bibinfo{number}{5} (\bibinfo{year}{2012}),
  \bibinfo{pages}{3250–3265}.
\newblock
\urldef\tempurl%
\url{https://doi.org/doi:10.1109/tit.2011.2182033}
\showDOI{\tempurl}


\bibitem[\protect\citeauthoryear{Wilhelmi, Barrachina{-}Mu{\~{n}}oz, Bellalta,
  Cano, Jonsson, and Neu}{Wilhelmi et~al\mbox{.}}{2019a}]%
        {sr_mab_2}
\bibfield{author}{\bibinfo{person}{Francesc Wilhelmi}, \bibinfo{person}{Sergio
  Barrachina{-}Mu{\~{n}}oz}, \bibinfo{person}{Boris Bellalta},
  \bibinfo{person}{Cristina Cano}, \bibinfo{person}{Anders Jonsson}, {and}
  \bibinfo{person}{Gergely Neu}.} \bibinfo{year}{2019}\natexlab{a}.
\newblock \showarticletitle{Potential and pitfalls of Multi-Armed Bandits for
  decentralized Spatial Reuse in WLANs}.
\newblock \bibinfo{journal}{\emph{Journal of Network and Computer
  Applications}}  \bibinfo{volume}{127} (\bibinfo{year}{2019}),
  \bibinfo{pages}{26--42}.
\newblock
\urldef\tempurl%
\url{https://doi.org/10.1016/j.jnca.2018.11.006}
\showDOI{\tempurl}


\bibitem[\protect\citeauthoryear{Wilhelmi, Barrachina-Muñoz, Cano, Selinis,
  and Bellalta}{Wilhelmi et~al\mbox{.}}{2021}]%
        {Wilhem2021}
\bibfield{author}{\bibinfo{person}{Francesc Wilhelmi}, \bibinfo{person}{Sergio
  Barrachina-Muñoz}, \bibinfo{person}{Cristina Cano}, \bibinfo{person}{Ioannis
  Selinis}, {and} \bibinfo{person}{Boris Bellalta}.}
  \bibinfo{year}{2021}\natexlab{}.
\newblock \showarticletitle{Spatial Reuse in IEEE 802.11ax WLANs}.
\newblock \bibinfo{journal}{\emph{Computer Communications}}
  \bibinfo{volume}{170} (\bibinfo{year}{2021}), \bibinfo{pages}{65--83}.
\newblock
\showISSN{0140-3664}
\urldef\tempurl%
\url{https://doi.org/10.1016/j.comcom.2021.01.028}
\showDOI{\tempurl}


\bibitem[\protect\citeauthoryear{Wilhelmi, Cano, Neu, Bellalta, Jonsson, and
  Barrachina-Muñoz}{Wilhelmi et~al\mbox{.}}{2019b}]%
        {sr_mab}
\bibfield{author}{\bibinfo{person}{Francesc Wilhelmi},
  \bibinfo{person}{Cristina Cano}, \bibinfo{person}{Gergely Neu},
  \bibinfo{person}{Boris Bellalta}, \bibinfo{person}{Anders Jonsson}, {and}
  \bibinfo{person}{Sergio Barrachina-Muñoz}.}
  \bibinfo{year}{2019}\natexlab{b}.
\newblock \showarticletitle{Collaborative Spatial Reuse in wireless networks
  via selfish Multi-Armed Bandits}.
\newblock \bibinfo{journal}{\emph{Ad Hoc Networks}}  \bibinfo{volume}{88}
  (\bibinfo{year}{2019}), \bibinfo{pages}{129--141}.
\newblock
\showISSN{1570-8705}
\urldef\tempurl%
\url{https://doi.org/10.1016/j.adhoc.2019.01.006}
\showDOI{\tempurl}


\bibitem[\protect\citeauthoryear{Williams and Rasmussen}{Williams and
  Rasmussen}{1995}]%
        {gp}
\bibfield{author}{\bibinfo{person}{Christopher K.~I. Williams} {and}
  \bibinfo{person}{Carl~Edward Rasmussen}.} \bibinfo{year}{1995}\natexlab{}.
\newblock \showarticletitle{Gaussian Processes for Regression}. In
  \bibinfo{booktitle}{\emph{Conference on Neural Information Processing Systems
  (NeurIPS'95)}}.
\newblock


\bibitem[\protect\citeauthoryear{Zhu, Guo, Yang, Conner, Roy, and Hazra}{Zhu
  et~al\mbox{.}}{2004}]%
        {Zhu2004}
\bibfield{author}{\bibinfo{person}{Jing Zhu}, \bibinfo{person}{Xingang Guo},
  \bibinfo{person}{L.~Lily Yang}, \bibinfo{person}{W.~Steven Conner},
  \bibinfo{person}{Sumit Roy}, {and} \bibinfo{person}{Mousumi~M. Hazra}.}
  \bibinfo{year}{2004}\natexlab{}.
\newblock \showarticletitle{Adapting physical carrier sensing to maximize
  spatial reuse in 802.11 mesh networks}.
\newblock \bibinfo{journal}{\emph{IEEE Wireless Communications and Networking
  Conference (WCNC'04).}} (\bibinfo{year}{2004}).
\newblock


\end{thebibliography}

\end{document}